\documentclass[sigplan, screen, authorversion]{acmart}

\usepackage{paralist}
\usepackage{breakurl}

\setcopyright{rightsretained}
\acmPrice{}
\acmDOI{10.1145/3359061.3362776}
\acmYear{2019}
\copyrightyear{2019}
\acmISBN{978-1-4503-6992-3/19/10}
\acmConference[SPLASH Companion '19]{Proceedings of the 2019 ACM SIGPLAN International Conference on Systems, Programming, Languages, and Applications: Software for Humanity}{October 20--25, 2019}{Athens, Greece}
\acmBooktitle{Proceedings of the 2019 ACM SIGPLAN International Conference on Systems, Programming, Languages, and Applications: Software for Humanity (SPLASH Companion '19), October 20--25, 2019, Athens, Greece}

\begin{document}
\title[Toward a Benchmark Repository for Software...]{Toward a Benchmark Repository for Software Maintenance Tool Evaluations with Humans}

\author{Mat\'u\v{s} Sul\'ir}
\orcid{0000-0003-2221-9225}
\affiliation{%
  \institution{Technical University of Ko\v{s}ice}
  \streetaddress{Letn\'a 9}
  \city{Ko\v{s}ice}
  \postcode{042 00}
  \country{Slovakia}
}
\email{matus.sulir@tuke.sk}

\begin{abstract}
To evaluate software maintenance techniques and tools in controlled experiments with human participants, researchers currently use projects and tasks selected on an ad-hoc basis. This can unrealistically favor their tool, and it makes the comparison of results difficult. We suggest a gradual creation of a benchmark repository with projects, tasks, and metadata relevant for human-based studies. In this paper, we discuss the requirements and challenges of such a repository, along with the steps which could lead to its construction.
\end{abstract}

%
%



\maketitle

\section{Introduction}

To evaluate a programming technique or tool, controlled experiments with human participants are often performed. The participants are divided into two groups, one using the tool/technique being evaluated and the other using the baseline. They are asked to perform the supplied maintenance tasks on a given project, such as fixing a bug or implementing a feature. Performance measures, e.g., the time to complete the tasks and their correctness, are collected and compared.

A crucial decision in this context is what project and tasks to select. Only 15\% of studies use real tasks from issue trackers, and even in this case, they are sometimes modified to fit the purpose \cite{Ko15practical}. The rest of the studies use made-up tasks, which can favor the evaluated tool, be unrealistic and thus decrease external validity: \textit{It would be surprising if the researcher could not come up with a single scenario where the new technique would prove itself somehow `better' than an existing technique \cite{Greenberg08usability}.} Furthermore, since everyone uses a different project and task for evaluation, it is difficult to compare techniques or perform data synthesis in systematic reviews. Although multiple software engineering artifact repositories exist \cite{Rodriguez12software,Boehme17where}, they are specialized or designed mainly for automated studies and not focused on experiments with humans. For these reasons, we advocate the creation of a repository of benchmarks for software maintenance tool evaluations with human participants. It should contain a set of projects and tasks which the researchers could use to evaluate tools, along with the results of already performed experiments.

\section{Projects}

Currently, researchers trying to select a project for use in an experiment can sift through a long list at websites such as GitHub, trying to find a project fulfilling all general criteria:
\begin{inparaenum}[1)]
\item it is an engineered software project -- not, e.g., a tutorial,
\item it can be successfully and effortlessly compiled from source,
\item the problem domain is general enough to be understandable by the participants,
\item it is relatively self-contained,
\item it does not require extensive manual configuration and setup,
\item it has automated tests with sufficient coverage.
\end{inparaenum}
The selected project must also fulfill experiment-specific criteria, such as:
\begin{inparaenum}[1)]
\item it is written in the given programming language,
\item it is either a library, a web application, or a mobile application, etc.,
\item the project either uses some specific technology/framework or does not use it,
\item it has a suitable size, depending on the kind of experiment being performed.
\end{inparaenum}
Based on our experience, this searching process can last even a few days.

We envision a website with a sample of projects fulfilling all of the general criteria. For each such project, specific criteria are listed, so the researcher can select a project according to the needs. For example, we can search for a medium-sized non-Android Java library. Note that selecting such specific criteria lowers the chance that multiple researchers will use the same project. Therefore, the researcher can also enter a simpler query, such as ``any Java project'' -- and the system will always return the same project. If we are not satisfied with the suggestion, the system can offer an alternative.

\section{Tasks}

There are two challenging criteria for a task:
\begin{inparaenum}[1)]
\item It should be representative of the tasks performed in practice. Since industrial tasks tend to be confidential, our best match is tasks from the issue trackers of open source projects, first automatically filtered and then manually curated.
\item The task should exercise the behavior tested in an experiment. For example, a bug in a single thread is useless for an experiment focused on multi-threaded debugging, even if it may be relevant with respect to the tasks occurring in practice.
\end{inparaenum}

To help researchers fulfill the second criterion, the benchmark repository would contain not only task descriptions copied from issue trackers, but also sets of task properties. The researcher could then filter the tasks according to the properties relevant for a given experiment. Now a difficult question arises -- what properties can a task have?

First, we can assign each task a course-grained category from a fixed set, such as a bug fix or feature addition. Then it is possible to divide the categories into more fine-grained ones, e.g., bugs into memory, concurrency bugs, etc. Care must be taken to find the right level of granularity, since too many categories would be impractical.

We can also ask: What activities should a developer perform in order to solve the task? By aggregating data from multiple developers working on the same task, we could obtain typical patterns of developers' behavior for the given task. What exactly is an activity, though?

It is possible to distinguish a fixed number of high-level activities, such as ``understanding'' or ``writing'' \cite{LaToza06maintaining}. Such activities can be obtained manually from screen recordings and think-aloud protocols, or automatically from IDE (integrated development environment) interaction traces \cite{Roehm12automatically}.

We can also include low-level actions recorded in the IDE and other windows during a task. Information about recorded events would be particularly useful when the tested tool focuses on certain actions in the user interface of an IDE. For example, if for a certain task, we found that the majority of developers spent at least 50\% of time in a debugger, it is a suitable task for the evaluation of a debugger enhancement.

Many researchers (starting with \cite{Sillito08asking}) studied what questions developers ask when programming. We could obtain a list of frequently asked questions for a particular task, based on the data from think-aloud protocols. Such data would help with the task selection, particularly if the evaluated tool is focused on answering a specific question.

At our envisioned benchmark website, the researcher will be able to use a specific query, such as ``a bug-fixing task when the text search was often performed and questions about variable changes were asked'' -- or a general query, e.g. ``any bug-fixing task''. More general queries increase the chance of finding a suitable task.

\section{Creation Steps}

Our vision can be realized on multiple levels, each useful even on its own:
\begin{inparaenum}[1)]
\item Structured demonstration: A group of researchers gathers to try their tools on the same project, comparing their results and experience. Such events already occurred \cite{Sim00structured}, but they are rare.
\item Contest: Besides demonstrations, the participants also compete based on a set of criteria (see, e.g., DocGen -- \url{http://dysdoc.github.io}).
\item Benchmark repository: In the information visualization community, such a repository was created using data from contests \cite{Plaisant08promoting}. Existing data from already performed experiments could also be added, in case they are sufficiently complete.
\end{inparaenum}

To illustrate how the envisioned repository could look like in the future, a simple static demo is at \url{http://sulir.github.com/humanbench}. A researcher will select a project and tasks based on criteria. Then it would be possible to download a container with the project, its dependencies, an IDE, and tasks descriptions. If no project/task fulfills the criteria, it should be possible to add a new one. Finally, the researchers should upload the results of their experiments to the repository, so it would be possible to see and compare them.

\begin{acks}
This work was supported by Project VEGA No. 1/0762/19 Interactive pattern-driven language development.
\end{acks}

\bibliographystyle{ACM-Reference-Format}
\bibliography{splash}

\end{document}